\documentclass{aastex631}
\usepackage{makecell}

\begin{document}

\title{Infrared Colors of Small Serendipitously-Found Asteroids in the UKIRT Hemisphere Survey}

\author[0009-0003-6656-7827]{Samantha G. Morrison}
\affiliation{Grinnell College, 1115 8th Ave., Grinnell, IA 50112, USA}
\affiliation{Northern Arizona University, 1900 S Knoles Dr., Flagstaff, AZ 86011, USA}

\author[0000-0001-6350-807X]{Ryder H. Strauss}
\affiliation{Northern Arizona University, 1900 S Knoles Dr., Flagstaff, AZ 86011, USA}

\author[0000-0003-4580-3790]{David E. Trilling}
\affiliation{Northern Arizona University, 1900 S Knoles Dr., Flagstaff, AZ 86011, USA}

\author[0000-0002-2601-6954]{Andy J. López-Oquendo}
\affiliation{Northern Arizona University, 1900 S Knoles Dr., Flagstaff, AZ 86011, USA}

\author[0000-0002-3858-1205]{Justice Bruursema}
\affiliation{United States Naval Observatory, Flagstaff Station, 10391 West Naval Observatory Rd., Flagstaff, AZ 86005, USA}

\author{Frederick J. Vrba}
\affiliation{United States Naval Observatory, Flagstaff Station, 10391 West Naval Observatory Rd., Flagstaff, AZ 86005, USA}

\correspondingauthor{Samantha Morrison}
\email{morrison2@grinnell.edu}

\begin{abstract}

The UKIRT Hemisphere Survey covers the northern sky in the infrared from 0--60~degrees declination. Current data releases include both J~and K~bands, with H-band data forthcoming. Here we present a novel pipeline to recover asteroids from this survey data. We recover 26,138~reliable observations, corresponding to 23,399~unique asteroids, from these public data. We measure J--K colors for 601 asteroids. Our survey extends about two magnitudes deeper than 2MASS. We find that our small inner main belt objects are less red than larger inner belt objects, perhaps because smaller asteroids are collisionally younger, with surfaces that have been less affected by space weathering. In the middle and outer main belt, we find our small asteroids to be redder than larger objects in their same orbits, possibly due to observational bias or a disproportionate population of very red objects among these smaller asteroids. Future work on this project includes extracting moving object measurements from H and Y band data when it becomes available.

\end{abstract}

\keywords{Asteroids (72), Sky surveys (1464), Near infrared astronomy (1093), Main belt asteroids (2036)}

\section{Introduction} \label{sec:intro}

Small bodies in the Solar System act as chemical and dynamical tracers of planetary system formation and evolution. Main belt asteroids (MBAs) --- a relatively accessible population, observationally --- have long been identified as indicators of Solar System history (e.g., \cite{demeo_2014}). To access and constrain the compositional gradient of the Solar System, the properties of very many asteroids must be measured. This would be an intractably large task if each asteroid were observed individually, but large-scale all-sky surveys produce large catalogs of asteroids that can provide powerful constraints. These archival searches can provide information for tens of thousands of asteroids and generally require no additional telescope time.

Previous work has shown substantial mixing of asteroid spectral types within dynamical groups (e.g., inner, middle, and outer main belt; \citet{demeo_2014}). This evidence supports a turbulent formation of the Solar System. Possible examples of turbulent formation models include the Grand Tack and the Nice Model, described in detail in \cite{walsh_low_2011} and \cite{Tsiganis}, respectively. 

Infrared observations are often the most powerful way to measure asteroid compositions, as many important biochemical markers (such as water) have absorption features in the infrared (e.g., \cite{binzel_2019}). However, large-scale all-sky infrared sky surveys are still a relatively recent opportunity as they require advanced technologies. Existing surveys, such as the Two-Micron All-Sky Survey (2MASS) and the UKIRT survey suite, were designed as Solar System surveys, not asteroid-specific data sets. Our archival search methodology mines existing data sets, extracting specialized asteroid results from more general all-sky survey data. \cite{sykes_2000} extracted 3804 asteroids from the Two-Micron All-Sky Survey (2MASS), which surveyed to the relatively bright limiting magnitude of $\sim$13.5, corresponding to an approximate object diameter of $\sim$6~km for main-belt asteroids. 2MASS was designed to study stars, not asteroids, so Sykes' asteroid recoveries are serendipitous within the data. Likewise, our recovered asteroids are serendipitous occurrences within the UKIRT Hemisphere Survey (UHS) catalog data.

The power-law shape of the asteroid size distribution implies that more sensitive surveys would include a substantially larger number of asteroids, producing a more powerful probe of the history of the Solar System. In service of this goal, \cite{popescu_2016} recovered 39,937 asteroids from VISTA survey data, with objects as small as $\sim$0.2~km. The UHS is complementary to VISTA and allows us to detect asteroids to $\sim$2~km size range. In this work, we recover and characterize 23,399 unique asteroids from the UHS data, which covers all of the Northern Hemisphere between declinations of 0° and 60°. Section \ref{sec:survey} reviews the relevant properties of the UHS. Section \ref{sec:methods} presents our novel pipeline to extract asteroids from UHS data. Section~\ref{sec:results} presents our catalog of extracted asteroids, and Section~\ref{sec:discussion} describes the science implications of our results. Finally, Section \ref{sec:conclusions} summarizes our discussion and provides suggestions for further work.

\section{Survey\label{sec:survey}}

\subsection{UKIRT, WFCAM, and the UHS\label{subsec:ukirt}}

UKIRT is a 3.8~m telescope located on Maunakea. UKIRT's wide-field camera (WFCAM) was designed to maximize survey speed with a large field of view of 0.8 square degrees. The camera utilizes four 2k $\times$ 2k infrared detectors and a pixel scale of 0.4 arcsec on the sky. The detectors are separated by 94\% of their active area, in a square pattern, which allows subsequent images to be slightly overlapped. The focal plane layout is detailed in Figure 1 of \cite{dye_2006}.

The UKIRT Hemisphere Survey (UHS) is an ongoing UKIRT program that surveys the Northern Hemisphere between decl. of 0° and 60° in the infrared Y, J, H, and K bands. The survey launched on 19 May 2012. The UHS was selected for this project primarily because of its wide sky coverage and relatively high sensitivity. While the UHS survey was initially designed as a Solar System survey, not specifically for finding asteroids, its extensive coverage means that many known main-belt asteroids are expected to fall within the survey's footprint. The J- and K-band surveys have been released, with H- and Y-band data forthcoming. Each data release contains 12,700 deg$^2$ of coverage, including J-/K-band imaging and source catalogs. The J-band data was made available to the public in August 2018, while the K-band is accessible to the public as of June 2023. The public releases cover regions of the sky not surveyed by the UKIRT Infrared Deep Sky Survey (UKIDSS) Large Area Survey, Galactic Plane Survey, or Galactic Cluster Survey regions. The data have a median 5$\sigma$ point source sensitivity of 19.6 mag in J and 18.4 mag in K (Vega), with a median full width at half-maximum of the point spread function across the data set of 0.75 arcsec in J (\cite{bruursema_2023}; Bruursema, et al. (2024), in preparation). These data allow us to measure the infrared colors of the asteroids we recover from the database. Our pipeline utilizes data from both releases to compare asteroid photometry in the J and K filters. For details of the coverage of the UKIRT surveys, see \cite{dye_2018}, especially Figure 2.

\subsection{WFCAM Data Products\label{subsec:data_products}}
The WFCAM Science Archive (WSA) website\footnote{\url{http://wsa.roe.ac.uk/}} has several publicly accessible data products, including image data, catalog data, and meta-data. Across the WSA data products, each image is assigned a unique ``multiframeID" for identification. This parameter is included in our final catalogs to facilitate cross-referencing with WFCAM data products. We utilize the meta-data to determine the central R.A,. decl., and observation time of each UHS image. We also utilize the image source catalogs, which detail the position and magnitude of each photometric source in the image. A photometric source is any point of light, most commonly stars. However, some of these sources are expected to be asteroids; our task is to determine which ones.

\section{Methods\label{sec:methods}}

We use the {\tt Ephem} and {\tt Orbit} functions from \textit{sbpy} \citep{mommert_2019}. The package makes use of the OpenOrb tool on the back-end while supplying more user-friendly methods for public access \citep{granvik_2009}. The accuracy and precision of \textit{sbpy}'s generated ephemerides are comparable precision to NASA's JPL Horizons System\footnote{\url{https://ssd.jpl.nasa.gov/horizons/app.html}} or MPC's Ephemeris Service\footnote{\url{https://www.minorplanetcenter.net/iau/MPEph/MPEph.html}}.

The core of our data processing pipeline is a Python program that cross-compares the positions of images of the UHS (in R.A., decl., and time) with the predicted positions of asteroids in the Minor Planet Center database. The pipeline works by taking a list of asteroid identifiers provided by the user. As it passes through the program, each asteroid will receive a flag indicating the result of the matching algorithm: matched or rejected, with the reason for rejection. To begin, the asteroids must pass a series of criteria before matching is attempted. The first check determines whether the specified asteroid intersects with the UHS survey footprint at any point during the survey's lifetime. This check is performed by calculating the decl. of each asteroid once a month for the duration of the survey using the \textit{sbpy} ephemerides tool. We then compute the percentage of those measurements that fall within the desired decl. range. If fewer than 1\% of the decl. measurements are in the range 0° $< \delta <$ 60°, the asteroid is flagged with a -1 in the full catalog and does not pass forward to the rest of the code. More than 99\% of asteroids pass this initial, very generous check. Although this first check does not eliminate many asteroids, it is computationally inexpensive, and for the asteroids it does reject, it prevents more expensive discernment down the line. For each asteroid that passes this initial check, we then generate a daily ephemeris for the duration of the survey. The daily ephemeris is then compared to the meta-data for the entire survey. Any image that falls within $\pm$ 12 hours in time and $\pm$ 0.5 degrees in both R.A. and decl. of the asteroid's ephemeris passes forward to the next stage of processing. If there are no image matches, the asteroid receives a flag of -2 and the process exits. If the asteroid has at least one possible image match, we calculate the position of the asteroid at the exact time of the image exposure. We confirm whether the asteroid appears within the particular location of any of the WFCAM's four panels at the time of the image exposure. If the asteroid is within the empty space between panels or outside the boundaries of the image, the asteroid/image pair receives a flag of -3. Asteroid observations that pass all of these tests are successful and assigned flags of 0. We repeat this process for all images along the asteroid ephemeris.

The program has now identified all images that contain the asteroid's predicted position and thus all asteroids that appear within the UHS. Next, for each image, we download the source detection table (unmerged source catalog) for the multiframeID from the WSA website utilizing an automated \textit{urllib} request. The unmerged detection tables contain individual passband detections that come from WFCAM images before multiple stationary source detections are merged. We use these catalog entries as the photometric measurements presented in this paper (in other words, we do not make any new measurements, but rather use catalog measurements reported by WSA). In contrast, the merged catalogs (source tables) are not suitable for asteroid archival recovery because the sources in the merged catalog are no longer tied to individual images. In the detection tables, each source is assigned a unique identifier known as its objectID. We search the image source catalog for the source (objectID) with the least absolute geometric offset from the asteroid's predicted position. The distance on the sky between asteroid (RA$_1$, DEC$_1$) and source (RA$_2$, DEC$_2$) is defined as:

\begin{center}
\begin{equation}
d = \sqrt{(DEC_2 - DEC_1)^2 + (RA_2 - RA_1)^2} .
\label{eq:dist}
\end{equation}
\end{center}

A more precise treatment of Equation \ref{eq:dist} would weight the R.A. component by $cos(DEC_2)$. Note that the survey coverage area has a maximum decl. of 60°, with $cos(60^{\circ}) = 0.5$. Moreover, the vast majority of our results are clustered near the ecliptic, with declinations well below the 60° limit (see Fig. \ref{fig:coverage} for the locations of successful matches). Therefore, we expect $cos(DEC_2)$ to be very close to 1 and have thus omitted this term. 

If there are no sources in the image within the prescribed limits, the asteroid/image pair is marked with the flag -4 and the program proceeds to the next image. The primary reason for this to occur would be due to a blank source catalog. A substantial number (approximately 20\%) of detection tables for valid multiframeIDs are empty, which would result in any asteroids supposedly in that image receiving the -4 flag. These multiframeIDs belong to observations that have been rejected due to seeing or cloud limits.

If an asteroid has been successfully paired with a best-fit source match (objectID) in an image, the pair is tagged with a flag of 0. The information about the match is in this case saved to the final catalog, as described in detail in Section~\ref{sec:results}. An asteroid may be found in multiple images, and an image may contain multiple asteroids. The catalog contains an entry for each unique recovery.

\begin{figure}
    \centering
    \includegraphics[width=.4\textwidth]{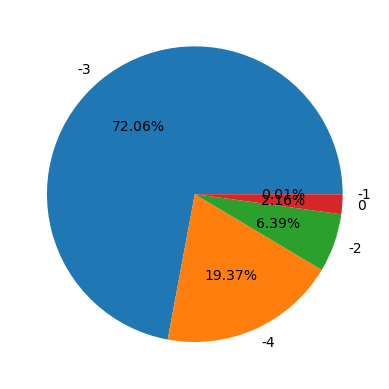}
    \caption{Prevalence of internal processing flags in preliminary asteroid match attempts. The flag meanings are detailed in Table \ref{tab:flags}. The majority of image/asteroid pair rejections are due to flag -3: asteroid found outside image's panels. This is likely due to the particularities of the WFCAM focal plane layout. The next most common flag is -4 due to blank source catalogs in the WFCAM Science Archive.}
    \label{fig:flags}
\end{figure}

A summary of the possible flags and their meanings is detailed in Table~\ref{tab:flags}. Their relative abundance in the data are shown in Figure~\ref{fig:flags}.

\begin{table}[!ht]
\caption{Internal Processing Flags - Exit Points for Main Pipeline}
\label{tab:flags}
\begin{center}
\begin{tabular}{ rl } 
\hline
Flag & Description \\
\hline
0 & Matched successfully \\
-1 & Asteroid is never in survey decl. range during course of survey\\
-2 & Asteroid is not near any specific images\\
-3 & Asteroid is outside image or in white space between panels\\
-4 & There are no sources in the image near the asteroid's predicted location (includes images with no source catalogs) \\
\hline
\end{tabular}
\end{center}
\end{table}

In total, we searched for all 623,827 numbered asteroids among all 927,233 public images. We recovered 142,319 potential asteroid matches, 26,138 of which have passed our internal vetting process (assigned flag=0) and are highly likely to be real asteroid recoveries. Our first data release containing the recovered matches is included as a machine-readable table. Descriptions of the columns are given in Table~\ref{tab:columns}.

\begin{table}[!ht]
\caption{Asteroid Catalog Columns}
\label{tab:columns}
\begin{center}
\begin{tabular}{ cll } 
\hline
Col. & Name & Description\\
\hline
 1 & astNum & Asteroid number \\ 
 2 & objID & Source identifier from WFCAM Science Archive catalog \\ 
 3 & multiframeID & Image identifier from WSA catalog\\ 
 4 & flag & One-digit code designating the result of our pipeline processing\footnote{See Table \ref{tab:flags}}\\
 5 & extNum & Extension number of the WFCAM frame \\
 6 & seqNum & Running number of the WFCAM detection\\
 7 & filterID & One-digit code designating the filter of the image\footnote{3=J, 5=K} \\
 8 & img\_MJD & Modified Julian Date of the image \\
 9 & astRa (deg) & Predicted R.A. of the asteroid (from {\em sbpy}) \\
 10 & srcRa (deg) & R.A. of the source in WSA catalog \\
 11 & raErr (arcsec) & Difference between astRa and srcRa \\
 12 & astDec (deg) & Predicted decl. of the asteroid (from {\em sbpy}) \\
 13 & srcDec (deg) & Decl. of the source in WSA catalog \\
 14 & decErr (arcsec) & Difference between astDec and srcDec \\
 15 & errSum (arcsec) & Square root of squares of raErr and decErr \\
 16 & aperMag3 & Calibrated and corrected aperture magnitude 3 (from WSA)\footnote{We use the ``aperture magnitude'' product from the standard WFCAM data processing, and refer to that measurement here as ``apparent magnitude.''} \\
 17 & aperMag3err & Error in calibrated aperture magnitude 3 (from WSA) \\
 18 & absMag & Asteroid absolute magnitude (from MPC) \\
 19 & G & Slope parameter (from MPC) \\
 20 & Vmag & Predicted visual magnitude (from MPC) \\
 21 & phAng (deg) & Phase angle (from MPC) \\
 22 & astHelDist (AU) & Heliocentric distance - instantaneous distance from the Sun (r) \\
 23 & astGeoDist (AU) & Geocentric distance - distance from the Earth (delta) \\
 24 & filtAbsMag & Calculated absolute magnitude in image filter \\
 \hline
\end{tabular}
\end{center}
\end{table}

\subsection{Data Processing\label{subsec:data_processing}}
To minimize the probability of false-positive matches, the post-processing algorithm rejects any unreliable entries. Entries are considered unreliable if the distance on the sky between the asteroid's predicted position and the source position exceeds 1.0 arcseconds. Finally, we eliminate any sources that have a reported photometric uncertainty $>$0.2~magnitudes. Between these criteria, approximately 82\% of match candidates are rejected, resulting in a final sample of 26,138~observations. Of these, 19,916~observations are in J, while 6,222 are in K. The observations represent 23,399~unique asteriods, some of which are multiply observed. We find 601~asteroids that have observations in both J and K, and thus have calculated J--K colors. There may exist correct matches among the rejected data, but more complex processing --- such as a stationary source removal --- is needed to more accurately distinguish between false positives and true matches.

\begin{figure}
\plottwo{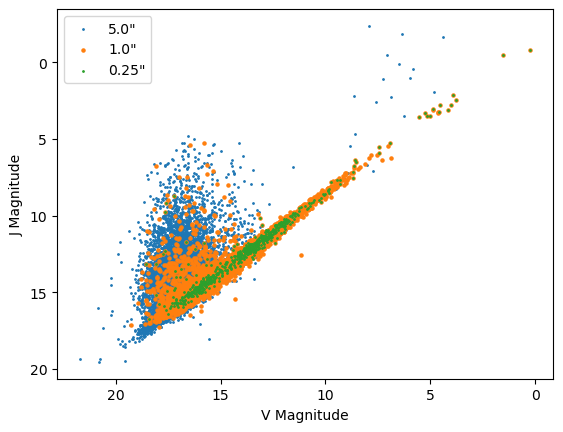}{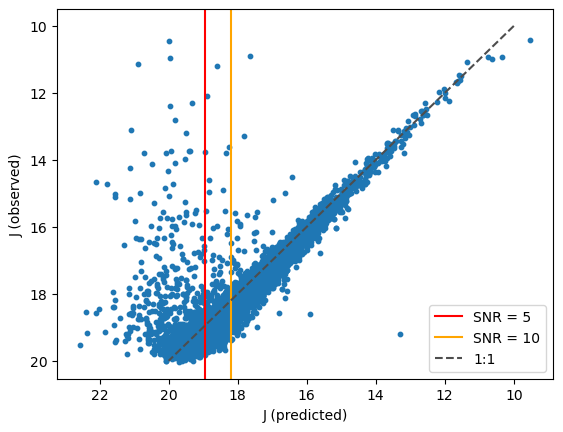}
\caption{Sample diagnostic plots for 19,916 asteroids detected in J band. Left: absolute magnitude in J vs. predicted absolute magnitude in V. The color indicates the varying search radius sizes (blue: 5.0" radius, orange: 1.0" radius, green: 0.25" radius). A larger match threshold creates more scatter away from the line, which are probably incorrect (false) matches. A smaller match threshold increases fidelity but decreases the overall size of our catalog. We choose a compromise value of 1~arcsec. Right: predicted J values (calculated from average V--J color of 1.54) vs. observed J values in WSA catalog. The vertical lines show the limiting magnitudes for SNR={5,10}.}
\label{fig:scatter}
\end{figure}

We retrieve from WSA apparent J or K magnitudes for each asteroid observation. We convert the apparent magnitudes to absolute magnitudes using Equation \ref{eq:absmag}, as described in detail in Section \ref{sec:results}; this calculation corrects for phase angle, as per the definition of absolute magnitude. We obtain predicted V magnitudes from MPC for each matched asteroid. We then derive J--K and V--J colors. Figure \ref{fig:scatter} shows the scatter resulting from different positional match (distance) thresholds of 5.0", 1.0", and 0.25" (blue, orange, and green points respectively). Figure \ref{fig:scatter}a shows the absolute J vs. absolute V of all 19,916 asteroids detected in the J band. Figure \ref{fig:scatter}b compares the J-band magnitudes from WFCAM data to predicted values. To obtain the predicted J values, the V--J colors of the asteroids with $<$0.25" match threshold are averaged to obtain the sample's average V--J color of 1.54. This value is then subtracted from MPC magnitudes, resulting in the predicted J values. Some false positives may be due to the asteroid’s position coinciding with a stationary source (such as a star), leading to a nonphysical V--J measurement. These sources would be eliminated by a stationary source removal algorithm by comparison with Gaia DR3 or other similar stellar catalog, as performed by others (e.g., \cite{sykes_2000}, \cite{popescu_2016}). Such a stationary source removal is beyond the scope of this work.

\subsection{Calculating Colors\label{subsec:calc_colors}}
\begin{figure}
\centering    
    \includegraphics[width=.4\textwidth]{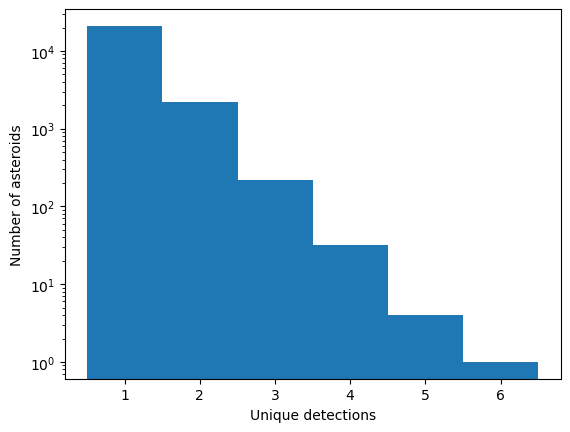}
\caption{The number of observations, in either filter, of the recovered asteroids. Most asteroids were found only once. The maximum number of observations for a single asteroid is six.
Only asteroids with 2~or more observations can possibly
have J--K colors derived.
}
\label{fig:repeat_obs}
\end{figure}

Because each observation in UKIRT is made in a single filter, the chance that a particular asteroid has observations in multiple filters is a function of observation repeats and some luck. To compute a J--K color of the asteroid, the survey must have captured the asteroid in both the J and K filter, requiring two separate crossings of the asteroid's ephemeris during times when the survey was utilizing the appropriate filters. The K-band data was obtained 5 years into the survey. Therefore, many asteroids detected in the J-band survey would have been below the declination limits of the K-band survey. Figure \ref{fig:repeat_obs} shows the repeated observations of asteroids recovered. No asteroid was recovered in more than six unique images, with most asteroids being recovered only once. As a result, our ability to measure variations within our data set is limited. Only 37 asteroids have more than three total measurements, making the data far too sparse for rotational lightcurve analysis. The effect of the sparse data is most significant in calculating J--K colors, which require at least two detections (one in each filter). 

Furthermore, since the J-band and K-band observations used to calculate J--K colors occur, in general, at different positions, the J--K color naively derived from the apparent magnitudes might be misleading. To mitigate the effects of differences in phase angle, heliocentric distance, etc., between the two measurements, the measured apparent magnitudes are converted to absolute magnitudes before calculating colors. The conversion is performed using the H and G magnitude system for asteroids:
\begin{equation}
    H = H(\alpha) + 2.5 \log((1-G) \phi_1 (\alpha) + G \phi_2 (\alpha)),
    \label{eq:absmag}
\end{equation}
where $H(\alpha)$ is the reduced magnitude (i.e. magnitude with the influence of distance removed), G is the slope parameter, 
\begin{equation}
    \phi_i (\alpha) = exp(-A_i (\tan{\frac{1}{2} \alpha})^{B_i} ),
\end{equation}
    $i$ = 1 or 2, $A_1$ = 3.33, $A_2$ = 1.87, $B_1$ = 0.63 and $B_2$ = 1.22 and $\alpha$ is the phase angle in degrees \citep{dymock_2007}. We use the HG system here as it is the most commonly used system (with G=0.15 assumed), as we do not have enough data for any given asteroid to re-derive properties via the HG12, HG12*, or other systems.

If an asteroid was recovered multiple times in the same filter, the calculated absolute magnitudes are averaged for each filter before the subtraction occurs. This process results in exactly one J--K color value for each asteroid that was observed in both filters. Finally, the colors are corrected to intrinsic (reflectance) colors by subtracting the J--K color of the Sun, which is 0.38~magnitudes in the Vega system \citep{willmer_2018}. Similarly, the V--J (Vega) color of the Sun is~1.17, for the UKIRT J filter and the Bessel Murphy V filter, used here as a proxy for MPC's ``V'' filter \citep{willmer_2018}.

\section{Results} \label{sec:results}

\subsection{Final Catalog\label{subsec:catalog}}

Our final catalog contains all asteroids that were matched with sources by the pipeline after the removal of false positives. The output columns are described in Table \ref{tab:columns}.

The final catalog contains a total of 26,138 recovered observations representing 23,399 unique asteroids. These include 31 near-Earth objects, 21,955 main-belt asteroids, and 1,413 objects beyond the main belt. The specific means by which we differentiated among these populations are described in Table~\ref{tab:families}.

\begin{table}[!ht]
\caption{Asteroid Families by Semi-Major Axis}
\label{tab:families}
\begin{center}
\begin{tabular}{ lll } 
\hline
Name & Semi-Major Axis Range (AU) & Number Recovered \\
\hline
Hungaria Group & 1.78 -- 2.06 & 26\\
Inner main belt & 2.06 -- 2.50 & 3871\\
Middle main belt & 2.50 -- 2.82 & 7782\\
Outer main belt & 2.82 -- 3.27 & 10302\\
Cybele Group & 3.27 -- 3.70 & 184\\
Hilda Group & 3.70 -- 4.20 & 228\\
Trojans & 5.10 -- 5.30 & 891\\
Outer Asteroids & 5.30 -- 30.1 & 87\\
TNOs & 30.1+ & 23\\
\hline
\end{tabular}
\end{center}
\end{table}

\begin{figure}
\centering    
    \includegraphics[width=.9\textwidth]{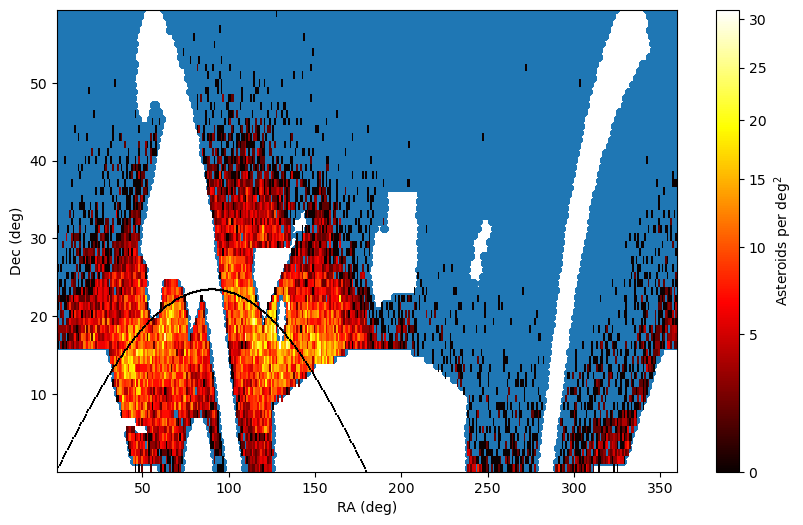}
    \caption{The positions of asteroid matches plotted over the footprint of the publicly-released UHS survey. White spaces designate sky coverage elsewhere in the UKIDSS series (Large Area Survey, Galactic Plane Survey, and/or Galactic Cluster Survey), which was excluded from the UHS public release. The matches are densely clustered around the ecliptic, as is expected for asteroid populations.}
    \label{fig:coverage}
\end{figure}
Our approximate limiting magnitudes are J$\sim$20, K$\sim$18, and V$\sim$20.

Figure \ref{fig:coverage} shows the position of the 26,138 asteroid recoveries over the pointings of the UHS survey. The color axis shows the number of asteroids per square degree. The majority of our asteroid recoveries lie along the ecliptic (shown in black). Most of the asteroids queried were previously known to reside close to the ecliptic.

\subsection{Survey Sensitivity\label{subsec:coverage}}

\begin{figure}
    \centering
    \begin{minipage}[b]{0.3\textwidth}
        \centering\includegraphics[width=\linewidth]{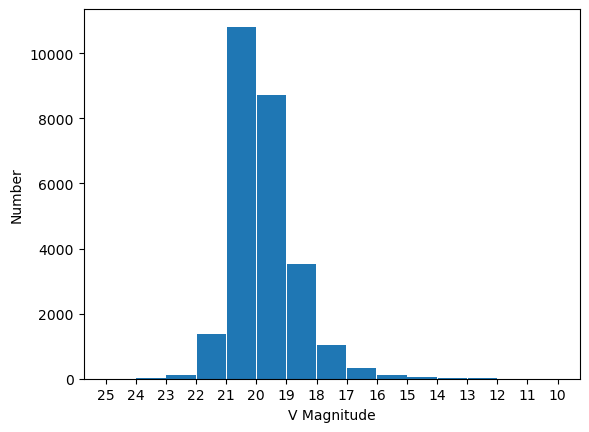} \\
    \end{minipage}%
    \hfill
    \begin{minipage}[b]{0.3\textwidth}        \centering\includegraphics[width=\linewidth]{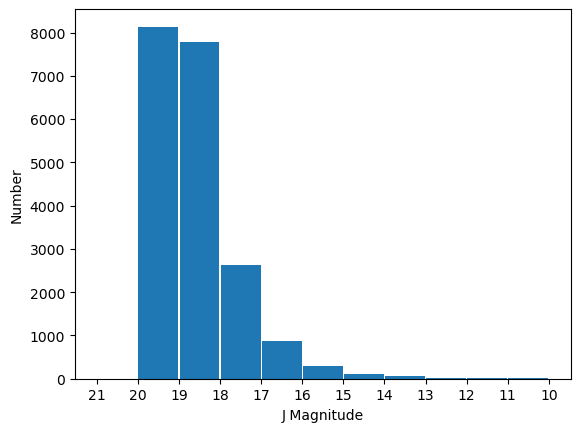} \\
    \end{minipage}%
    \hfill
    \begin{minipage}[b]{0.3\textwidth}
        \centering\includegraphics[width=\linewidth]{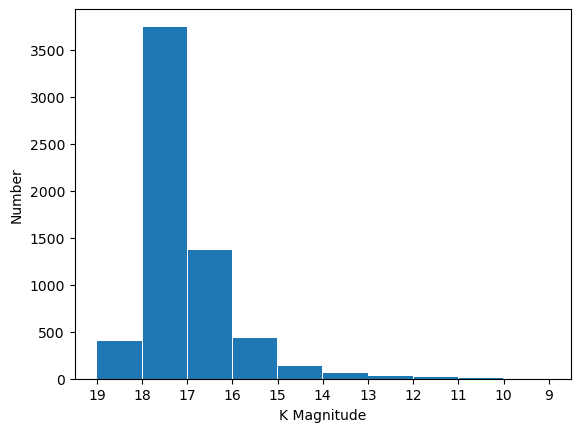} \\
    \end{minipage}
    \caption{Histograms of the distribution of asteroid magnitudes in the V (left), J (center), and K (right) filters. There are 19,916 asteroid recoveries in J and 6,222 total asteroid recoveries in K. The V magnitudes are for all 26,138 recoveries and are predicted by the MPC as queried by asteroid number. The approximate limiting magnitudes of our detections are J$\sim$20, K$\sim$18, and V$\sim$20. A few objects have V--J $>$ 4 and are likely false positives. This is discussed at length in Section \ref{sec:discussion}.}
    \label{fig:distributions}
  
\end{figure}

Figure \ref{fig:distributions} shows the distribution of our recovered asteroids in the V, J, and K filters. Note that while J and K values are observationally obtained from UKIRT data, the V magnitudes are those predicted by the MPC entries of the identified asteroid. Our approximate
limiting magnitudes are J$\sim$20 and K$\sim$18. Sykes' recovery is complete to approximately 17th magnitude; our data shows an improvement of two magnitudes.

\subsection{J--K Colors\label{subsec:colors}}
We calculated J--K colors for 601 asteroids (about 2.6\% of recovered asteroids) that have UKIRT measurements in both the J and K filters. The asteroids have an overall average intrinsic J--K color of 0.17 $\pm$ 0.40, with inner, middle, and outer main belt asteroids having average J--K colors of 0.06 $\pm$ 0.21, 0.14 $\pm$ 0.35, and 0.19 $\pm$ 0.43 respectively. The vast majority of these asteroids do not have previously reported infrared colors: 503 do not have colors reported in \cite{sykes_2000} or \cite{popescu_2016} or spectra reported in \cite{tholen_1984} or \cite{bus_2002}. 

\begin{figure}
    \centering
    \plottwo{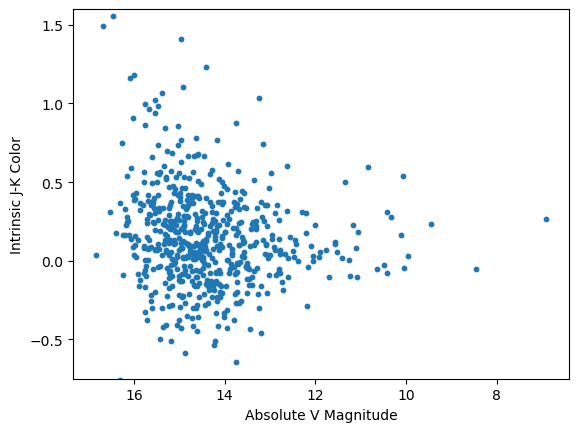}{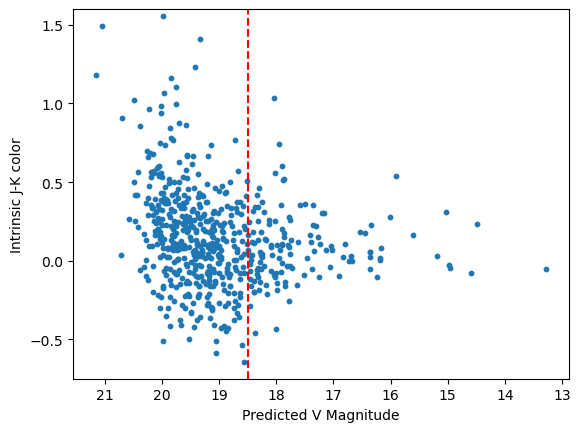}
    \caption{Dependence of J--K color on absolute (left) and apparent (right) V magnitude. The vertical line on the right panel shows the approximate limiting magnitude of \cite{sykes_2000}. The majority of our objects lie to the left of this line, indicating that our data represent a novel magnitude range that generally corresponds to smaller objects.}
    \label{fig:jk_sizes}
\end{figure}

Furthermore, we investigated the dependence of J--K color on asteroid absolute magnitude (as a proxy for asteroid size) and V magnitude. Figure \ref{fig:jk_sizes} shows these distributions. The figures suggest more variation in J--K color in smaller objects, although more data are needed to confirm whether this trend is due to the intrinsic properties of asteroids, or simply an artifact of our relatively small sample size. This trend makes statistical sense: if the distribution has a constant standard deviation as a function of size, we would expect to see more spread in the smaller objects where we have more total objects. Since our measurements were obtained near the detection limits of the UHS survey, our smaller objects in particular incur large photometric uncertainties. These uncertainties may inform the shape of the distribution. Therefore, the spread of our data may also be explained by intrinsic photometric errors in the UHS survey. 

\begin{figure}
    \centering
    \plottwo{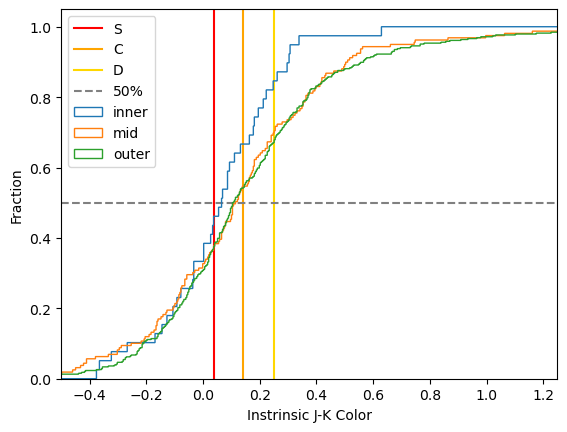}{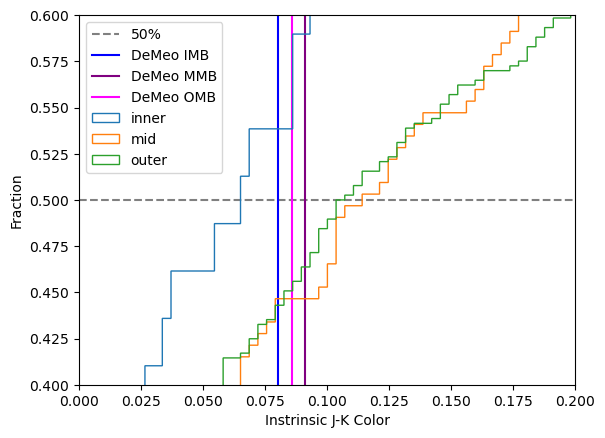}
    \caption{Cumulative distribution of J--K colors for
inner, middle, and outer main belt asteroids. The steepest slope belongs to the inner belt, showing that the inner belt is less dominated by very red objects than the other two regions. Left: The colors of S-, C-, and D-type asteroids are marked, along with a 50\% line for reference. There are several possible causes for the middle and outer belt presenting redder colors than the inner belt, as described in Section~\ref{sec:discussion}. Right: Same cumulative distributions as the left panel, but zoomed in. Here instead of showing mean colors for S, C, and D types, we show the mean calculated color for inner, middle, and outer main belt asteroids from the smallest size range (5--20~km) in \cite{demeo_2014}. In the inner main belt, our results are bluer than DeMeo predicts. In the outer and middle main belts, our results are less red than
the DeMeo result, while our middle and outer belt samples are more red. This result is discussed in detail in Section~\ref{sec:discussion}.}
    \label{fig:cumulative_hist}
\end{figure}

We convolved the UKIRT bandpasses with canonical spectra from \cite{demeo_2009} to derive intrinsic J--K colors for each taxonomic type. DeMeo determined taxonomic distributions for different populations for various size ranges (Figure 4 of \cite{demeo_2014}). We calculated the mean color for the 5--20~km size range --- their smallest size bin --- by weighting each taxa's color by its fractional abundance. We present the comparison between these DeMeo colors for each taxonomic type and our observed colors in Figure \ref{fig:cumulative_hist}b. Note that the DeMeo colors are not phase-corrected. We assume that the value of G (Equation \ref{eq:absmag}) is wavelength independent, meaning that our correction to absolute magnitude (0° phase) does not introduce a color. To further examine whether there could be an artificially introduced color difference between the DeMeo non-phase corrected colors and our phase-corrected colors, we examine \cite{sanchez_2012}, who show (in their Figure 5) the effect of increasing phase angle on the spectral slope of an ordinary chondrite (an analog to S-type asteroids). They find that the difference in the spectral slope from phase angles 13° to 30° is negligible. Considering that main-belt asteroids are never observed at phase angles $>$ 30° due to the geometry of the Solar System, and because \cite{sanchez_2012}  consider S-types as end-members with the greatest slope change, the effect should be even less pronounced for other spectral types in our data. Therefore, we can reasonably compare the colors between the DeMeo un-phase-corrected spectra and our phase-corrected results without worrying about phase-induced color differences, and phase reddening cannot explain the reddening trend present in our data. The inner main belt line crosses the 50\% mark to the left of the calculated mean color for inner main belt asteroids in the 5--20~km size range. Thus, our inner main belt objects, which are smaller than the DeMeo objects, are on average less red than the DeMeo objects. This could be due to space weathering, as described in Section \ref{sec:discussion}. On the other hand, the middle and outer main belt lines cross the 50\% mark to the right of the calculated mean color for their respective regions. In these two regions, our smaller objects are on average redder than the DeMeo objects. This could be due to an overabundance of very red objects in the middle and outer main belts at smaller sizes, as described in Section \ref{sec:discussion}. 

Figure \ref{fig:cumulative_hist} shows the cumulative distributions of the measured colors of inner, middle, and outer main belt asteroids.
We find a steeper slope for inner main belt asteroids than for middle or outer main belt asteroids (Figure~\ref{fig:cumulative_hist}a). This result implies that the inner main belt has fewer (very) red objects than either the middle or outer belt. According to our calculated colors of asteroid types in the infrared, S-type asteroids are less red than their C-type companions. Therefore our results show the presence of more S-complex asteroids in the inner main belt than in the middle or outer main belt. This finding is consistent with the findings of \cite{demeo_2014}.

\section{Discussion} \label{sec:discussion}

\subsection{Our Sample and Previous Work}\label{subsec:sample}

In total, we have recovered 26,138 observations of 23,399 unique asteroids, the vast majority of which are in the main belt. We have measured J--K colors for 601 of these asteroids. We hope to increase our number of detections with additional data from future UHS data releases (e.g., H-band, Y-band, repeat J-band, and repeat K-band observations).

Our population is an excellent companion to Popescu's (2016) recovery of Southern Hemisphere asteroids from the VISTA-VHS survey (MOVIS). The MOVIS catalogs contain a total of 39,947 objects, including 38,428 main-belt asteroids \citep{popescu_2016}. 73 of our asteroids are also observed in the MOVIS catalog. Further work should combine our photometric data with MOVIS data to obtain even more J--K colors from our existing recovered asteroids. VHS continues surveying the sky. Any asteroids exiting our survey range are entering the VHS survey footprint and have the potential to be observed by VHS.

\begin{figure}
\plottwo{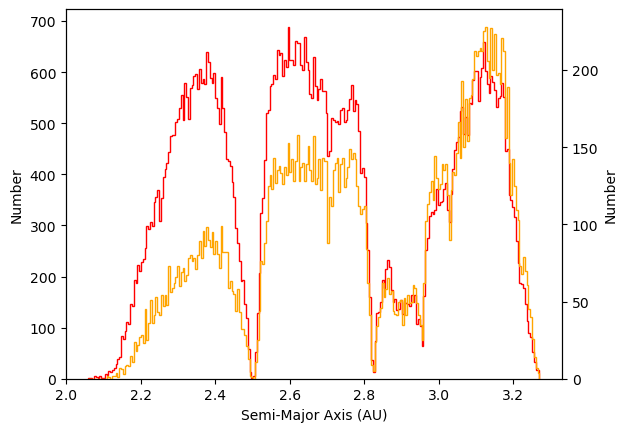}{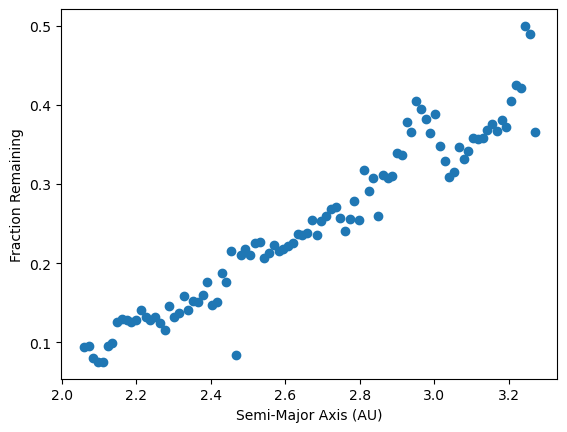}
\caption{Comparison of recovery rate before and after introducing a maximum of 1.0\arcsec\ between predicted location and source. Left: The red line (left y-axis) shows detected asteroids with no match threshold, and the orange line (right y-axis) shows detections with a 1.0" threshold. The peaks are set to the same height to demonstrate the recovery rate in each bin. Right: Fraction of matches retained after imposing the 1.0\arcsec\ limitation, as a function of semi-major axis. The recovery rate is a function of semi-major axis, with inner main belt asteroids affected more strongly. A possible explanation is that the faster speeds in the inner belt create more mismatches.}
\label{fig:compare_mb}
\end{figure}

\subsection{Potential Biases}\label{subsec:potential_biases}

The size difference between the inner and middle/outer regions may be due to an observational bias. We can see smaller objects in the inner belt because they are closer, whereas the same number of photons from the outer belt corresponds to a bigger object. If it is true that the outer belt asteroids are red due to their older surfaces, whereas the inner belt asteroids are less red (with younger surfaces) then our data hint at the timescale of space weathering, as described below.

Our match threshold of 1.0\arcsec\ introduces a potential bias in our final catalog. Asteroids with smaller semi-major axes disproportionately failed this threshold test (Figure \ref{fig:compare_mb}). Asteroids in the inner belt have greater sky velocities, so a slight mismatch between known and true orbital elements leads to an on-sky positional mismatch that is a function of orbital semi-major axis: larger for inner belt objects and smaller for outer belt objects. This would contribute to overall poorer matches for closer asteroids, as observed in Figure \ref{fig:compare_mb}. To remedy this bias, we intend to implement a stationary source removal. After the stationary source removal, each image source catalog will contain many fewer points, meaning we can safely expand the search radius without incurring as many false positive matches. We will still need to control for false positives (for example, by comparing observed and predicted values for each point in the same band), but can be much more generous and robust in our matching search radius.

Our data may contain a detection limit bias toward recovering redder objects. According to the UKIRT Time Exposure Calculator,\footnote{\url{https://ukirt.ifa.hawaii.edu/web/cgi/ITC/itc.pl}} the limiting magnitudes for a signal-to-noise ratio of 5, given the survey's 10-second exposure time, are 18.95 magnitude in J and 17.61 magnitudes in K. These limits correspond to a J--K color of 1.34. WFCAM is less sensitive in the K band, due to fundamental performance issues in the infrared. This, combined with the fact that asteroids are fainter in K band (primarily due to the Sun being fainter in K band), explains the reduced recovery in K as compared to J. This in turn requires objects to be redder (brighter in K) than average to be recovered near the detection limits. 

To illustrate the point, imagine the recovery of an object with J=18.95 exactly. If that object happens to be very red (J--K$>$1.34, such that K$<$17.61), the object may be observed in the K filter. However, if the object were less red (J--K$<$1.34), then the UHS would be unable to recover a K measurement for the object, as the object is too faint in K. Therefore, the unequal limiting magnitudes in J and K may introduce a bias toward redder objects, especially among the smallest objects. This trend is evident in Figure \ref{fig:jk_absmag_regions}. The smallest objects in the outer and middle main belt are, on average, redder (higher on the graph) than the overall average colors for their regions. For the faintest objects (small, located in the further-out middle and outer belts), we may be missing the bluer objects due to our detection limits, meaning the true population may not be quite as red as it seems.

The reddest data points could also be due to confusion of background stellar sources. Figure \ref{fig:scatter}a shows that at our chosen 1.0\arcsec\ error radius, there are still a few objects with V--J$>$3. These objects are likely background stars that have been confused with asteroid observations. This can occur either because the asteroid was too faint to be detected in the image or because the asteroid was not in its predicted position. Asteroids that are too faint will not result in an entry in the WSA detection catalog. Our program will still identify the nearest source to the asteroid's predicted position, which will in this case be a nearby star. Similarly, if the asteroid's true position in the image is different than its predicted position and there happens to be a stellar source that lies closer to the predicted position, we will measure the star rather than the asteroid.

\subsection{Colors of Our Objects}\label{subsec:our_colors}

Figure \ref{fig:cumulative_hist} shows the color distributions for the inner, middle, and outer belt. To test whether there is a significant color difference across these regions, we performed a Kolmogorov-Smirnov test between each possible pair of zones (i.e., inner/middle, middle/outer, and inner/outer). The test shows no statistically significant difference in colors in any of these pairs, with p-values of 0.097, 0.765, and 0.071 respectively. Low p-values indicate that two populations are not drawn from the same parent population --- i.e., low p-values indicate different populations --- so these relatively large p-values suggest no statistically significant difference between these pairs of populations, with the middle/outer pairing appearing to be particularly similar and the inner belt appearing to be different, though not at a significant level. We note that these results are still based on a relatively small sample of 601 asteroids. In future data releases (such as H-band or additional K-band data) a much larger catalog might allow us to detect subtle differences among these groups.

\begin{figure}
\centering    
    \plotone{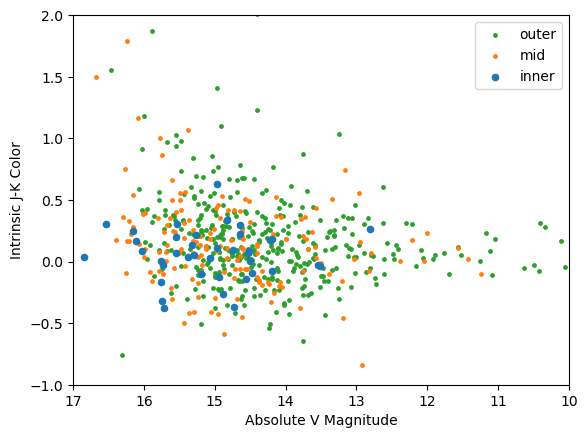}
    \caption{Dependence of J--K color on absolute magnitude by region. Recovered inner main belt objects are, on average, both smaller and less red than their counterparts in the middle and outer main belt.}
    \label{fig:jk_absmag_regions}
\end{figure}

\cite{deelia_2007} show that the collisional timescale for small bodies has a very steep dependence on size in the approximate regime of our objects (around $\sim$2~km). Their Figure 10 shows that the collisional timescale for 3~km bodies in the inner belt is 100 Myr and the collisional timescale for 5~km bodies in the outer belt is $>$1 billion years. Thus, small bodies would have younger surfaces, and larger asteroids would have older surfaces. Our observed asteroids are smaller than those in the DeMeo sample. In the case of the inner main belt, the difference between our result and DeMeo's may be due to space weathering. If the color differences described above are due to space weathering, then our data suggest that the space weathering timescale in the belt is a few hundred Myr. A nuance here is that we would expect space weathering to be faster in the inner belt and slower in the outer belt, and that recovered inner main belt objects are smaller on average than outer and middle main belt objects (Figure \ref{fig:jk_absmag_regions}).

\begin{figure}
    \centering
    \plotone{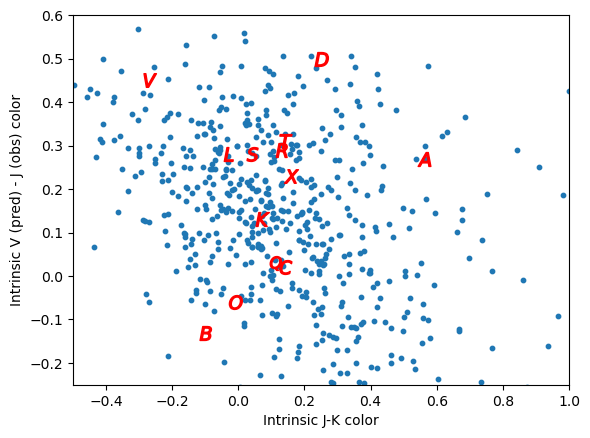}
    \caption{VJK color-color diagram of asteroids recovered. Each symbol represents the typical position in VJK color space of the specified type of asteroid. The canonical asteroid colors are derived from spectra from \cite{demeo_2014}, as described in the text. The blue points show the 601~asteroids in this survey with J--K colors. Our sample consists primarily of C- and S-type asteroids. K-, L-, Q-, R-, T-, and X-types fall near the S- and C-types and are difficult to distinguish in this color-color space.} 
    \label{fig:color_color}
\end{figure}

\subsection{Taxa of Our Asteroids}\label{subsec:our_taxa}

Although UHS data are currently only available in the J and K bands, we can simulate a color-color diagram utilizing predicted V magnitude measurements from MPC. We can then begin taxonomical categorization of our recovered asteroids. The mapping of typical asteroid types over our recovered data is shown in Figure \ref{fig:color_color}. The colors of the DeMeo types are derived by convolving V, J, and K filters with the canonical spectra from \cite{demeo_2009}. The plot shows that most of our objects are S-like or C/Q-like, while few are A, B, D, O, or V types. This is to be expected given the relative abundance of these types in the main belt. However, the presence of a few extremely red objects (J--K $>$ 0.8) explains the overall trend of reddening in the middle and outer main belt. The location of K-type asteroids may be misleading --- we would not expect as many K-type asteroids as there are data points near the K label. It may be the case that K-type asteroids simply lie at the intersection between the S cloud and C cloud. 

Figure \ref{fig:color_color} shows the colors of our catalog asteroids, together with mean colors for all major taxonomic types. Notably, this figure shows that we have recovered very few A-, B-, D-, O-, or V-type asteroids. In some cases, this may be explained by observational biases. For example, D-types have relatively low albedos \citep{Thomas_2011}, and may be disproportionately excluded by reflected-light surveys such as this one. However other types (e.g., A-types) are not significantly affected by this low albedo bias, so our low recovery rate of these types suggests that these types are not common among small objects. A- and O-types, in particular, are relatively rare in the main belt \citep{demeo_2014}, so it may not be surprising to see few of them in our sample. Our data also show the presence of some asteroids with J--K colors redder than any convolved type. This may be caused by variations in colors of taxonomic types or uncertainties in our measurements (with limits of 0.2~mag, some of these very red points might marginally be consistent with the colors of A-type asteroids, though red J--K and blue V--J is not the color of any known taxonomic type). Further data are needed to better constrain the transitions between particular taxonomic classes in VJK color space. Once the H- and Y-band data are made available by future UHS data releases, we will be able to refine our analysis significantly and compare directly with previous results \cite[e.g.,][]{sykes_2000,popescu_2016}.

\section{Conclusions and Further Work} \label{sec:conclusions}
We have presented our processing steps to extract serendipitously-observed asteroids from the UKIRT Hemisphere Survey (UHS). We recover 26,138 total observations of 23,399 unique asteroids, the vast majority of which are in the main belt. The inner main belt asteroids are less red, while the middle and outer main belt asteroids are redder than previous literature suggests. To explain the former, we suggest space weathering, as our objects are smaller --- with, presumably, younger surfaces --- than those reported in previous work. For the latter effect, we explore observational or detection limit biases or a larger-than-expected population of very red objects among very small asteroids.

The UHS has additional data forthcoming: a future ($\sim$1.5~years) H-band release and a future simultaneously-obtained Y and second epoch J-band release ($\sim$3~years). The H-band sensitivity, combined with J--H colors of most asteroid types, implies that the next data release will significantly increase the number of serendipitously measured UHS asteroids. The future YJ release will again increase the number of asteroids that can be identified and classified.

The \textit{Euclid} mission, launched July 2023, will observe many moving  objects: \cite{carry_2018} estimates that \textit{Euclid} has the potential to observe about 150,000 Solar System objects in VIS, \textit{Y}, \textit{J}, and \textit{H} down to $V_{AB}$ $\sim$~24.5. Euclid surveys at ecliptic latitudes $>$15 degrees (compare to Figure \ref{fig:coverage}), and therefore the Euclid Solar System object catalog will be highly complementary to the result we present here.

The Legacy Survey of Space and Time (LSST) will soon provide an inventory of Solar System that will include an estimated 6 million small-body sources, more than a factor of five greater than are currently available \citep{lsst}. The LSST data will produce ugrizy colors for nearly all of these asteroids. Near infrared colors of asteroids are highly complementary to, and often more diagnostic than, optical colors. We assume that almost all of our asteroids will eventually appear in the LSST-derived catalogs, so our objects can be used to calibrate LSST-derived taxonomic classifications.

In the future we can increase our catalog size by processing archival data from other large UKIRT programs including the UKIDSS Large Area Survey, Galactic Plane Survey, and Galactic Cluster Survey. As described above, further post-processing efforts will prioritize reducing false positive detection, especially through distinguishing between background source confusion and nearby (fast-moving) asteroids. A stationary source removal tool will be useful for this task. 

\section{Acknowledgments} \label{sec:acknowledgments}

This work is supported in part by the Northern Arizona University Research Experiences for Undergraduates program under National Science Foundation Grant 1950901.

This work is supported in part by the National Aeronautics and Space Administration under Grant No.\ 80NSSC20K0670 issued through the SSO Planetary Astronomy Program.

Taxonomic type results presented in this work were determined, in whole or in part, using a Bus-DeMeo Taxonomy Classification Web tool by Stephen M. Slivan, developed at MIT with the support of National
Science Foundation Grant 0506716 and NASA Grant NAG5-12355.

This research has made use of the VizieR catalogue access tool, CDS, Strasbourg, France. The original description of the VizieR service was published in A\&AS 143, 23.

\software{Astropy \citep{astropy:2013, astropy:2018, astropy:2022}, OpenOrb (https://github.com/oorb/oorb), \textit{sbpy} (https://sbpy.org/)}

\bibliography{citations}{}
\bibliographystyle{aasjournal}

\end{document}